\documentclass[prc,preprint,showpacs]{revtex4}
\usepackage{graphicx,dcolumn,array,bm,amsmath,amssymb}

\def\be{\begin{equation}}
\def\ee{\end{equation}}
\def\bea{\begin{eqnarray}}
\def\eea{\end{eqnarray}}
\def\beas{\begin{eqnarray*}}
\def\eeas{\end{eqnarray*}}

\begin{document}

\title{New method for calculating shell correction}

\author  {P. Salamon}
\affiliation{
Institute of Nuclear Research of the Hungarian  Academy of Sciences,
H-4001 Debrecen, P. O. Box 51, Hungary,\\
University of Debrecen, Faculty of Informatics, H-4010 Debrecen, P. O. Box 12,
Hungary}
\author{A. T. Kruppa}
\affiliation{
Institute of Nuclear Research of the Hungarian  Academy of Sciences,
H-4001 Debrecen, P. O. Box 51, Hungary}
\author{T. Vertse}
\affiliation{
Institute of Nuclear Research of the Hungarian  Academy of Sciences,
H-4001 Debrecen, P. O. Box 51, Hungary, \\
University of Debrecen, Faculty of Informatics, H-4010 Debrecen, P. O. Box 12,
Hungary}
\date{\today}

\begin{abstract}
A new method is presented for calculation of the shell correction with the
inclusion of the continuum part of the spectrum.
The smoothing function used has a finite energy range in contrast to the Gaussian
shape  of the Strutinski method.  The new 
method is specially useful for light nuclei where the generalized Strutinski 
procedure can not be applied.
\end{abstract}



\maketitle

\section{Introduction}
Nuclei being far from the bottom of the stability valley are studied extensively
at the experimental facilities with radioactive beams.
 One of the fruit of these type of research
is the production  of the light exotic nuclei. Let us refer to e.g.
a recently identified new double magic nucleus
the $^{24}O$ \cite{[kan09]} at the neutron drip line.
 The exact location of the particle drip lines limits the region for these
studies and it is intensively
investigated both by experimental and theoretical methods.
 Theoretical prediction of the drip lines 
is based on mass (binding energy) calculations since
particle separation energies can be easily deduced.

There are two important theoretical frameworks for global mass calculations.
 Microscopic HF or HFB calculations with sophisticated effective
density dependent interactions are very successful in this field. 
In the best HFB  mass formula so far \cite{atk1}  the rms error is 674 keV
\cite{Lun03}. 
In earlier HF calculations \cite{atk2,atk3} this number was somewhat larger,
namely 805 $keV$ and 822 $keV$ \cite{Lun03}. 
In order to achieve this improved fit a  new parameterization of the effective 
nucleon-nucleon interaction has been introduced and the pairing interaction 
was treated differently than in the earlier calculations.

Surprisingly a more simple alternative procedure in the framework
of the so called macroscopic microscopic (MM) 
formalism can compete with the microscopic calculations in the calculation 
of the binding energies.
The rms error in the MM calculation is 676 keV. We may say that the quality 
of the microscopic and MM methods are the same. 
Despite the almost identical global fits however the microscopic and MM methods 
show considerable differences when the neutron drip line is 
approached \cite{Lun03}. 

The key quantity of the MM calculations is the shell
correction.
The concept of the shell correction was suggested 
long time ago by Strutinski \cite{[str67],[str68]} and it is 
still in use.
E.g. in a recent global mass calculation \cite{Bha09} the basic ingredient of the 
shell correction method the
smoothed single particle density is calculated in a semi-classical way by the
Wigner--Kirkwood
expansion. 
The other elements of the Strutinski method was not altered.

Since the invention of the shell correction there were several refinements of 
the original method. Besides the original energy averaging, a smoothing
in the  particle number space was
introduced \cite{Ton82, Pom04}. 
Even a combination of the two averaging spaces was considered
\cite{Dia04}. The particle mean field, the simple harmonic oscillator or Nilsson potential 
was replaced in the calculations by more realistic
phenomenological forms  in which the spectrum has a continuum beside the
discrete single particle levels. The treatment of the single particle level density due to the
continuum was a long standing problem \cite{[Ro72],[Naz94]} but an elegant solution was finally reached
\cite{Kru98,[Ve98]}.

A large part of the uncertainty due to the
proper choice of the technical parameters of the smoothing method has been
removed by introduction of the generalized Strutinski procedure\cite{[Ve98],[Ve00]}, which
made it possible to calculate reliable shell correction values for medium and
heavy nuclei, where the smoothed level density has a long region with linear
energy dependence. As it will be discussed in Sec. IV., for lighter nuclei the 
length of the linear region
is reduced due to the reduction of the number of the occupied shells and the 
increase of the shell gap. For light nuclei the lower
and upper ends of the spectrum distort linearity, therefore the method is not 
appropriate for light nuclei.

The main goal of this work is to develop a new method which is free from this
 limitation and is applicable for the whole nuclear chart, even in the vicinity
 of the two drip lines. We are solving this problem by introducing a finite range smoothing instead of
 the infinite range Gaussian smoothing used in the Strutinski method.
 
The paper is organized as follows. In Sec. II. we recapitulate the 
formalism of the calculation of the shell
 correction. In Sec. III. we describe the standard Strutinski method with
 the plateau condition. In Sec. IV. we do the same with the
 generalized Strutinski procedure, what we want to replace in this work. 
 In Sec. V. we
 describe the new method with finite range smoothing in details. In Sec. VI
we apply the new method for
 several nuclei and calculate shell corrections for neutrons and protons. Finally
 in Sec. VII. we end with
 the main conclusions of the paper.

\section{Calculation of the binding energy by using the shell correction.}

 The binding energy of an atomic
nucleus composed of $A=N+Z$ nucleons ($N$ neutrons and $Z$ protons) $B(N,Z)$ can
be calculated in the microscopic-macroscopic model  (MM)  as
\be
\label{binding}
B(N,Z)= E_{macr}(N,Z) + \delta E(N,Z)~,
\ee
where $E_{macr}(N,Z)$ is the binding energy calculated in the macroscopic model
(e.g. liquid drop or droplet model) and $\delta E(N,Z)$ is the shell correction.
While $E_{macr}(N,Z)$ is a smooth function of the number of nucleons, 
the shell correction takes care of the shell fluctuations of the binding energy
which is missing from the macroscopic model. Shell fluctuations are present in
any microscopic model. E.g. the shell correction can be calculated
from single particle energies of  self-consistent Hartree-Fock  and  relativistic mean filed calculations 
\cite{[Be03],[Re06]}. In Ref. \cite{[Re06]} shell correction calculated on 
the
single particle energies was used to generate a smooth energy from the result
of these microscopic calculation and the typical phenomenological 
parameterization of the {\em microscopically calculated} macroscopic energy 
terms were analyzed.

  In the present work  we use the simplest i.e. the independent particle
shell model to generate the single particle energies in a phenomenological
nuclear potential for the sake of simplicity only, since the smoothing procedure
could be tested equally well on the result of this simple model.
In this model we treat neutrons and protons separately. In
this case the shell correction
\be
\label{nzshell}
\delta E(N,Z)=\sum_{\tau=\nu,\pi}\delta E_\tau (N_\tau)=\delta E(N)+\delta E(Z)
\ee
is the sum of the shell corrections $\delta E_\tau(N_\tau)$ calculated for 
neutrons: $\tau=\nu$ 
with $N_\nu=N$  and
for protons $\tau=\pi$
with $N_\pi=Z$ . In what follows we shall discuss the calculation
of the shell correction  $\delta E_\tau(N_\tau)$ for a sort
of nucleons only.

 The shell correction can be
estimated as the difference of the shell model binding energy
$E_{sp}^{\tau}$ and its smoothed counterpart
$\tilde {E}^\tau$ calculated also in the shell model.
\be
\label{shellc}
\delta E_\tau=E_{sp}^\tau-\tilde {E}^\tau~.
\ee
Here the shell model binding energy 
\begin{equation}
E_{sp}^\tau =
\sum_{j=1}^{N_\tau} E_j^{\tau} 
\end{equation}
is a sum of the
single particle energies $E_j^{\tau}$  of the lowest
energy orbits, from $E_1^\tau$  until the Fermi-level.
In the sum above we can take into account the $n_i$-fold degeneracies of the shell model
orbits and use only the different single particle energies denoted by $e_i^{\tau}$ 
\begin{equation}
E_{sp}^\tau =
\sum_{i} n_i~ e_i^{\tau}~. 
\end{equation}

The key quantity of the MM model is the smoothed energy  $\tilde {E}^\tau~$ 
therefore, we have to
give a unique definition for calculating it unambiguously. 
If we have the bound single particle energies: $e_i^\tau$, the density of the 
bound nuclear levels is
\begin{equation}\label{discreteleveldens}
g_{\rm d}^\tau(E)=\sum_i n_i~ \delta(E-e_i^\tau).
\end{equation}
The particle number as a functions of the energy $E$ of the single nucleon considered 
is an integral of the
level density in Eq.(\ref{discreteleveldens}), i.e. it is equal to the following step function:
\be
\label{partnumb}
n^\tau(E)=\int_{-\infty}^E  g_{\rm d}^\tau(e) de=\sum_i n_i~ \Theta(E-e_i^\tau)~,
\ee
where $\Theta(x)$ is a Heaviside function of the form:
\begin{equation}
\label{heavi}
\Theta(x)=\left\{
\begin{array}{rl}
0 &\textrm{, if } x~<~0\\
1&\textrm{, if } x~\geq~ 0~.
\end{array}
\right.
\end{equation}

Since in the smoothing procedure  we treat neutrons and protons on the same
footing, we can drop the $\tau$ index for a moment. ( We shell include it 
later again when it is needed to avoid ambiguity.)
 We can calculate the smoothed level density $\tilde{g}(E)$ 
from the level density in Eq.(\ref{discreteleveldens}) by folding it with a 
properly selected smoothing function: $f_p(x)$. The smoothing function
 spreads the energy of a discrete level over a certain energy range
 characterized by the smoothing range parameter $\gamma$. Therefore, the smoothed level
 density is
\begin{equation}\label{ggsmooth}
\tilde{g}(E) = \frac{1}{\gamma}
\int_{-\infty}^{+\infty} g(e)~
f_p\left(\frac{e-E}{\gamma}\right) de~.
\end{equation}
The smoothing function in Eq.(\ref{ggsmooth}) is usually a product of a weight function
$w(x)$ and a polynomial $h_p(x)$ of degree $p$
\begin{equation}\label{smothfunc}
f_p(x)=w(x)~h_p(x).
\end{equation}
The later is called as {\em curvature correction polynomial}.
Since the smoothing function $f_p(x)=f_p(-x)$ is an even 
function of $x$, for an even weight function $w(x)$ the polynomial $h_p(x)$ 
should also be even and the coefficients 
of the odd    terms in it
 should be equal to zero.
Therefore, the curvature correction polynomial has the form:
\begin{equation}\label{curvcorr}
 h_p(x)=\sum_{i=0,2,...,p} c_i x^i~ .
\end{equation}
The $c_i$ coefficients of the curvature correction polynomial $h_p(x)$ 
 are determined from the so called {\em self-consistency condition} 
 \cite{[Bun72]},
 which requires
that the smoothing should reproduce the original function if it is a polynomial
$g_n(x)$ with degree $n \le p+1$:
\be\label{selfcon}
g_n(x)=\int_{-\infty}^{+\infty} g_n(x')~f_p(x-x')dx'~.
\ee
 
We calculate the smoothed energy
by using the smoothed level density in Eq.(\ref{ggsmooth}) :
\begin{equation}\label{Esmooth}
\tilde{E}=\int_{-\infty}^{\tilde\lambda}
\epsilon\, \tilde{g}(\epsilon) d\epsilon~.
\end{equation}
The smoothed Fermi-level $\tilde{\lambda}$ is calculated from the               
condition that the number of neutrons and protons, i.e. the particle number is 
given:
\begin{equation}\label{ltilde}
N = \int_{-\infty}^{\tilde{\lambda}} \tilde{g}(\epsilon)d\epsilon~.
\end{equation}
The smoothed Fermi-level $\tilde{\lambda}$ is different from the  Fermi-level
$\lambda$ because the level density has been modified by the smoothing.

\section{Standard Strutinski method with plateau condition}

Strutinski used a  smoothing function with a Gaussian
a weight function
\begin{equation}\label{weight}
w(x)= \frac{1}{\sqrt{\pi}}\exp(-x^2)~, 
\end{equation}
and it can be shown that
the curvature correction polynomials for a weight function of Gaussian shape are the associated 
Laguerre-polynomials
\be\label{lag}
h_p(x)=L_{p/2}^{1/2}(x^2)~.
\ee
Therefore, in the standard Strutinski method the smoothing function is
\be\label{smf}
f_p(x)=\frac{1}{\sqrt{\pi}}\exp(-x^2)L_{p/2}^{1/2}(x^2)~.
\ee
For nuclei lying on the bottom of the stability valley the 
single particle potential can be approximated by a simple harmonic oscillator (h.o.) form.
For a nucleus with mass number $A$ the distance of consecutive shells can be expressed by the well 
known rule \cite{[Mo57]}
\be
\label{hejtav}
\hbar \Omega_0= 41~ A^{-1/3}~~ [MeV]~.
\ee
Shell structure of this simple h.o. model is modified by the presence of the
spin-orbit interaction and also by the non-spherical shape of deformed nuclei
but the quantity in Eq.(\ref{hejtav}) is still serves as a reasonably good measure for the
shell structure.
  An attractive feature of the h.o. potential is that  the shell correction $\delta E(\gamma,p)$
as a function of the smoothing range $\gamma$
  shows a wide plateau in which the
\begin{equation}\label{plato}
\frac{\partial
\delta E(\gamma,p)}{\partial \gamma}=  0
\end{equation}
  {\em plateau condition} is fulfilled. More precisely, the fulfillment of the 
{\em plateau condition} is valid  if at the same time the values belonging to 
the plateau  are practically independent of the $p$ value used.
It was observed that the {\em plateau condition} is fulfilled for h.o. 
potential. Since $\gamma$ and $p$ are  technical parameters of the smoothing 
procedure and they have no physical meaning, it is natural
to expect that the definition of the smoothed quantities should not depend
strongly on
these values.
Therefore, the shell correction calculated for the h.o. potential is well defined.
This nice feature of the h.o. potential is related to the fact that this
potential has only bound states (even at high positive energy values).
For potentials which are similar to the harmonic oscillator potential
e.g. the Nilsson potential we can always find regions for
 $\gamma$ where the 
{\em plateau condition} is fulfilled \cite{[Bra73],[Ro72]}. 
Since these potentials
have only bound states (infinitely many) and no continuum the
 ending of the bound states does not spoil the picture.
 
\section{Generalized Strutinski procedure for spectra with continuum}

However a more realistic single particle potential has a discrete spectrum 
with finite number of bound states
 $e_i<0$ and a continuum of scattering states with $E>0$  energy. The full
 level density in this case is a sum of the level densities of the discrete
 states $g_{\rm d}(E)$ and that of the
 scattering states $g_{\rm c}(E)$ forming the continuum 
\begin{equation}\label{e3}
g(E) = g_{\rm d}(E) + g_{\rm c}(E).
\end{equation}

Now the smooth level density has to be calculated again with the prescription of
Eq.(\ref{ggsmooth}).
It was realized  by Brack and Pauli\cite{[Bra73]} that for this case the plateau condition
can not be satisfied since the $\delta E(\gamma,p)$ curves, what we call
{\em plateau curves} do not have wide plateaus, where Eq.(\ref{plato}) is
fulfilled.  They searched for the minima $\delta E (\gamma_p,p)$ of the plateau curves for each
$p$ values and
 introduced the concept of
{\em local plateau condition}. At the minima i.e. at $\gamma=\gamma_p$  Eq.(\ref{plato}) is certainly 
satisfied. 
 An additional
requirement of the {\em local plateau condition} is the approximate 
$p$-independence of the $\delta E (\gamma_p,p)$
values, which is satisfied if the variation of the $\delta E (\gamma_p,p)$
values are small.

\begin{figure}[tbp]
\includegraphics[width=15.0cm]{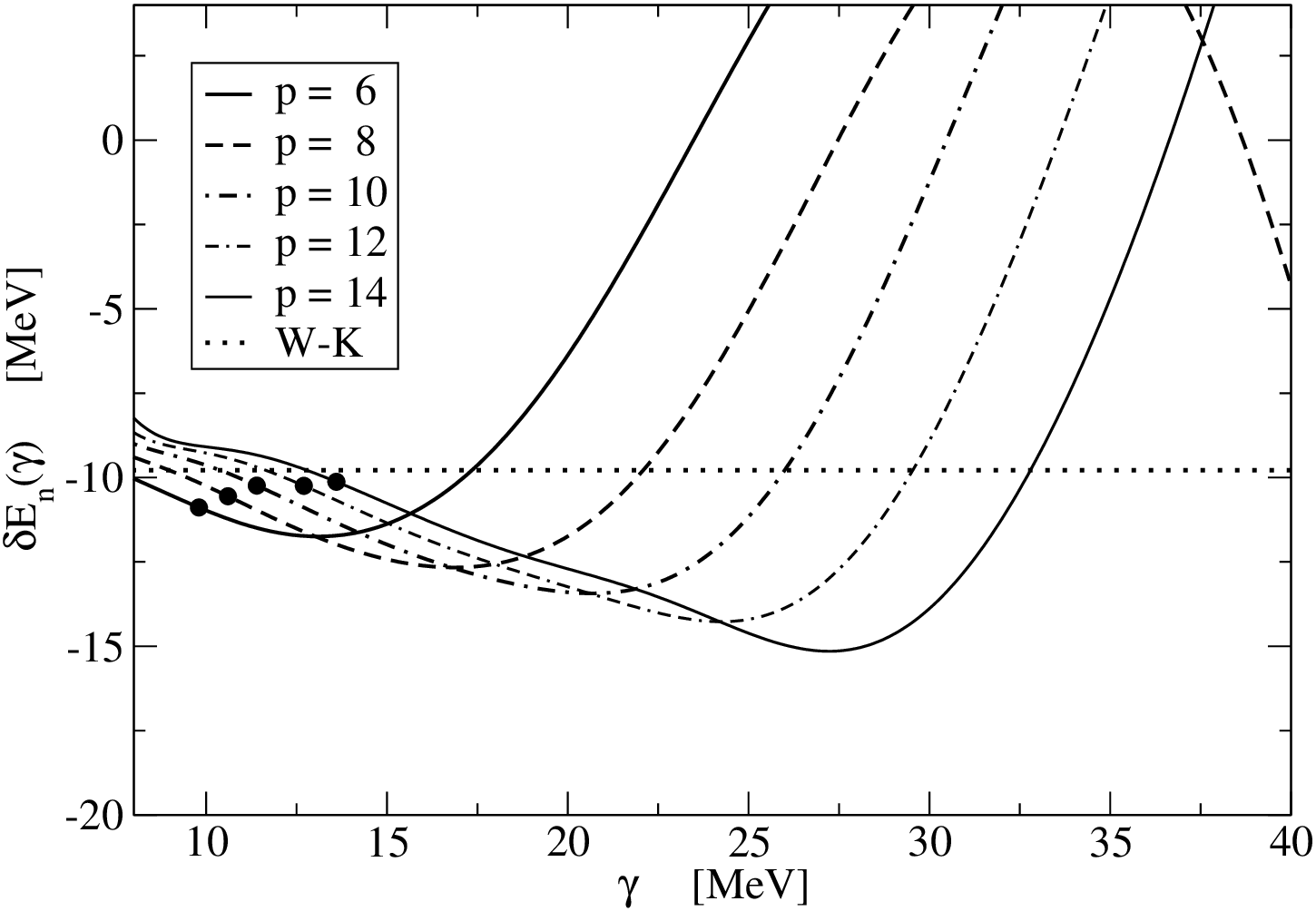}
\caption{Neutron shell correction $\delta E_n(\gamma,p)$ for the nucleus $^{146}Gd$ as a function of the smoothing range
$\gamma$ calculated for $p=6,..,14$ by using the Gaussian 
weight function for the smoothing
functions $f_p$. 
Filled circles on the different curves denote the $(\gamma_p,\delta
E_n(\gamma_p))$ points, where $\gamma_p$ values belong to the minima of the
function in Eq.(\ref{khi2}) and the $\delta E_n(\gamma_p,p)$ values are the results of the generalized 
Strutinski procedure.
Dotted horizontal line shows the value of
the semi-classical value $\delta E_{sc}=E_{sc}-E_{sp}^n$.}
\label{gdg}
\end{figure}

It was shown in Ref.\cite{[Ve98]} that sometimes even the 
{\em local plateau condition} might not be fulfilled and the smoothing procedure of the standard Strutinski
method might not able to furnish us with well defined smoothed energy. A typical
nucleus for which the local plateau condition fails if the continuum part of the 
spectrum is taken into account is the $^{146}Gd$, as one can see 
in Fig. \ref{gdg}.  Although
one can find minima for each plateau curves, the shell correction values at 
these minima
vary too much (even an approximate $p$-independence is not hold). Therefore
it is not surprising that the
$\delta E(\gamma_p,p)$ values deviate considerably from the semi-classical value.

In order to cure this difficulty in the work \cite{[Ve98]} a
 {\em modified plateau condition} was suggested.
 In the {\em modified plateau condition} the plateau condition in Eq.(\ref{plato})
 is replaced by the requirement that in a certain energy region the smoothed
 level density should be fitted well by a straight line. 
 
The shell correction $\delta E (\gamma_p,p)$ for a given $p$ 
should be calculated with those $\gamma_p$ value for which
the smoothed level density can be fitted best by a linear function:
$y(E)=a E + b$ in a certain energy range: $[e_l,e_u]$. 
So we should find the minimum of the function in the variable $\gamma$ for
each $p$ value
\begin{equation}\label{khi2}
\chi ^2(\gamma,p)=\sum_{i=1}^{n_u}\Big [{\tilde
g}(q_i,\gamma,p)-y(q_i)\Big ]^2.
\end{equation}
Here $q_i$ for $i=1,..,n_u$ is a mesh
of the energy interval $[e_l,e_u]$ used, and $\gamma_p$ is the value where the
function $\chi ^2$ has its minimum at a given $p$-value.
To get rid of the shell
fluctuations the length of the interval has to be larger than the estimated shell
gap
\begin{equation}\label{sav}
e_u -  e_l = 1.5~\hbar\Omega_o~.
\end{equation}
Having selected the proper $\gamma_p$ value for a set of $p$ values between
 $p_{min}=6$ and $p_{max}=14$,  the mean value and the variation
 of the corresponding $\delta E (\gamma_p,p)$ values have to be calculated as
 follow:
\begin{equation}\label{avrage}
\delta {E} = \frac{2}{(p_{max}-p_{min}+2)}\sum_{p=p_{min},p_{min}+2,...,p_{max}}
~
\delta  {E}(\gamma_p,p)~,
\end{equation}
\be
\label{vari}
\sigma=\sqrt{\frac{2}{(p_{max}-p_{min}+2)}\sum_{p=p_{min},p_{min}+2,...,p_{max}}
~(\delta
{E}(\gamma_p,p)-\delta {E} )^2}~.
\ee
Since in Ref.\cite{[Ve98]} this variation was reasonably small for most of the 
nuclei, the mean in Eq.(\ref{avrage}) was used to 
define the shell correction and the variation in Eq.(\ref{vari}) was considered as an
uncertainty of the method.
The procedure described above was called as a {\em generalized Strutinski procedure}.

\begin{figure}[tbp]
\includegraphics[width=15.0cm]{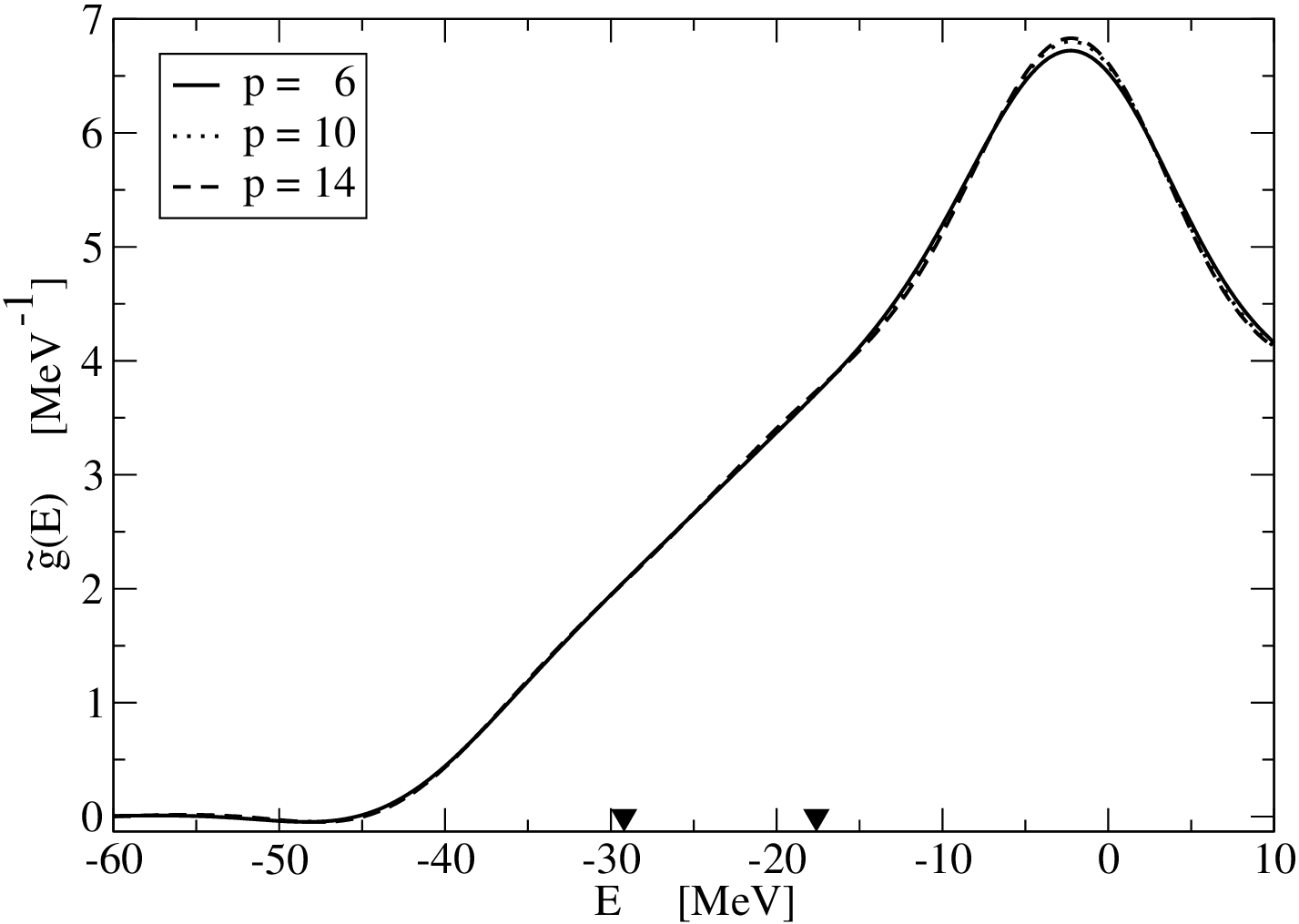}
\caption{Energy dependence of the smoothed level densities calculated in the generalized 
Strutinski procedure for $p=6,10,14$ by using a Gaussian weight function for the smoothing
functions $f_p$ for the nucleus $^{146}Gd$ . The lower and upper ends of the 
interval $[e_l,e_u]$ in which the
condition of the best linear fit is applied are shown  by triangles
on the $E-$ axis.}
\label{smold}
\end{figure}

In order to illustrate the use of the {\em modified plateau condition} we
present the smoothed level densities for the $^{146}Gd$ nucleus in Fig.
\ref{smold}. The lower and upper ends of the energy interval in which the best linear
fit of the $\tilde g(E)$ is required are shown by filled triangles on the
$E$-axis. Practically no $p$-dependence of the $\tilde g(E)$ curves can be
observed in the $[e_l,e_u]$ interval where $\tilde g(E)$ is apparently behaves
as a linear
function of $E$. Some $p$-dependence can only observed
at around $E\approx -10$ MeV being a bit above the $\tilde {\lambda}$ value and
 at higher energy in the $E=0$ MeV region which has no influence on the
 shell correction. 
The large bump of the smoothed level density around  $E=0$ MeV is the effect of
the higher end of the spectrum. In the positive part of the spectrum only a few
neutron resonance contribute to the level density and their effect is smoothed
by the smoothing parameters which are the abscissas of the filled circles in 
Fig.\ref{gdg}. These $\gamma_p$
values are between $10-15$ MeV, therefore the end effect is spread well below
the threshold. The effect of the lower end is less pronounced but can be seen
at $E<-35$ MeV. Here the derivative of $\tilde g(E)$ with respect to $E$ 
changes and at $E<-45$ MeV $\tilde g(E)$ goes below zero for a while. The main
feature of the $\tilde g(E)$ is that the linearity required in Eq.(\ref{gdg}) 
holds only in a certain distance from the lower and upper ends of the spectrum.

  In Fig.\ref{gdg} the filled circles on
the different $p$ curves show the $(\gamma_p,\delta E_n(\gamma_p,p))$ points
where the $\gamma_p$ values are those where the function in Eq.(\ref{khi2}) has its
minimum. One can see from the circles that these  shell correction values have 
much smaller  variation ($\sigma$) than the shell correction values at the
minima of the curves. Moreover the mean of the $\delta E_n(\gamma_p,p)$ values
denoted by circles is in good agreement with
the dotted line showing the semi-classical value.
In the work \cite{[Ve98]}  it was found that this situation is quite typical and
 the {\em generalized Strutinski procedure} gave similar values to 
   the result of the semi-classical averaging based on the Wigner--Kirkwood 
   expansion \cite{[Bra73],[Bha71],[Jen73],[Jen75],[Jen75a],[Jen75b],[Bra97]} 
   in those cases in which the later could be applied.
Moreover the {\em generalized Strutinski procedure} gave similar results 
to that of the standard one for all cases where the {\em plateau condition} is fulfilled.
But it gave a well defined value for the smoothed energy even in cases like
$^{146}Gd$ where we can not really speak about plateau. 
    
It turned out only later, in the work \cite{[Ve00]}
where the {\em generalized Strutinski procedure} was used for deformed nuclei,
that the function in Eq.(\ref{khi2})
might have more than one minimum in $\gamma$. 
 It was concluded in that  the minimum at the
smaller $\gamma$ value should be selected.

An uncertainty of the generalized Strutinski smoothing procedure is that the results are slightly
depend on the position of the $[e_l,e_u]$ energy interval used.
For medium and heavy nuclei the uncertainty of the generalized Strutinski procedure was always 
below $250$ keV.
To get this small variation, the energy interval $[e_l,e_u]$ was adjusted to the smoothed
Fermi-level, and the upper end of the energy interval  was
$e_u=\tilde\lambda-\hbar\Omega_0$.
If the interval was shifted up to have $e_u=\tilde\lambda$ and the length
was kept the same as in Eq.(\ref{sav}) a variation of the shell correction
by around 400\,keV was observed. This uncertainty was still reasonably small and it was
comparable to the typical deviation from the semi-classical result.

 The dependence on the position of the interval become stronger for light 
nuclei.
If the mass number $A$ is reduced, the distance of the shells estimated in
Eq.(\ref{hejtav}) increases and the length of the interval in Eq.(\ref{sav}) also
increases. We should use larger and larger $\gamma$ values for smoothing the 
shell fluctuations. On the other hand
the region in which $\tilde{g}(E)$ is linear becomes shorter and shorter
because the effect of the lower end shifts higher and that of the higher end
shifts lower. Therefore for small $A$ there is not enough space
where the required linear region could develop. 
The linearity of $\tilde{g}(E)$ function is spoiled by the end effects.
This explains why the {\em generalized Strutinski procedure} breaks down for light nuclei. 

Therefore, in this work our goal is to find a new smoothing procedure which is less 
sensitive to the end effects, but it still keeps the advantages of the generalized Strutinski
procedure i.e. the shell correction is practically independent of the $p$ values ($\sigma$
is small). An additional requirement is that $\tilde{E}$ resulted by the new 
procedure should not be too different from the result of the semi-classical 
procedure (Wigner--Kirkwood method)  if the later approach can be applied. 

\section{New smoothing procedure}

A disadvantage of the smoothing procedures used so far is that the Gaussian
weight function $w(x)$ used  has an infinite range, therefore, the effect of an energy
$e_i$ is smeared to the whole energy axis from $-\infty$ to $\infty$. Therefore,
the effect of the lower and upper ends of the spectrum influences the whole
region of the smoothed level density and also the shell correction $\delta{E}$.
In this work we try to reduce the end effects in these quantities by using weight
functions which have only a finite range.
One possible candidate for a weight function with finite range is a shape
\be
\label{finitew}
w(x)=\left\{
\begin{array}{rl}
k e^{-\frac{1}{1-x^2}}&\textrm{, if } |x|~<1\\
0&\textrm{, if } |x|~\geq 1.
\end{array}
\right.
\ee
The value of the normalization constant $k$ should be chosen from the condition that
\be
1=\int_{-1}^{+1} w(x)~ dx~.
\ee
One advantage of the form in Eq.(\ref{finitew}) is that all derivative of that
function are continuous at $|x|=1$, so the weight function continues smoothly 
 to the regions where it is equal to zero.
The effect of the smoothing with this form is localized to the $x\in [-1,1]$
interval.
In order to use the new smoothing function we have to 
recalculate the curvature correction polynomials $h_p(x)$ 
in Eq.(\ref{curvcorr}) for the new weight function (in Eq.(\ref{finitew})).
The recalculated polynomials $h_p(x)$ will be different from the one in Eq.(\ref{lag})
and they should satisfy the self-consistency condition in Eq.(\ref{selfcon}),
with the finite-range weight function.
As it was shown in Ref.\cite{[Bun72]}, the coefficients $c_i$ of the curvature correction polynomials in
 Eq.(\ref{curvcorr}) are solutions of the system of linear equations:
\be
\label{line}
\sum_{i=0}^p c_i a_{i+j}=\delta_{j,0}~~~~~~~~~0\leq j\leq p~,
\ee 
where the coefficients $a_l$ are the integrals:
\be
a_l=\int_{-1}^{1} w(x) x^l~ dx~.
\ee
The integration is over the interval where the weight function $w(x)$ is
different from zero.

\begin{figure}[tbp]
\includegraphics[width=15.0cm]{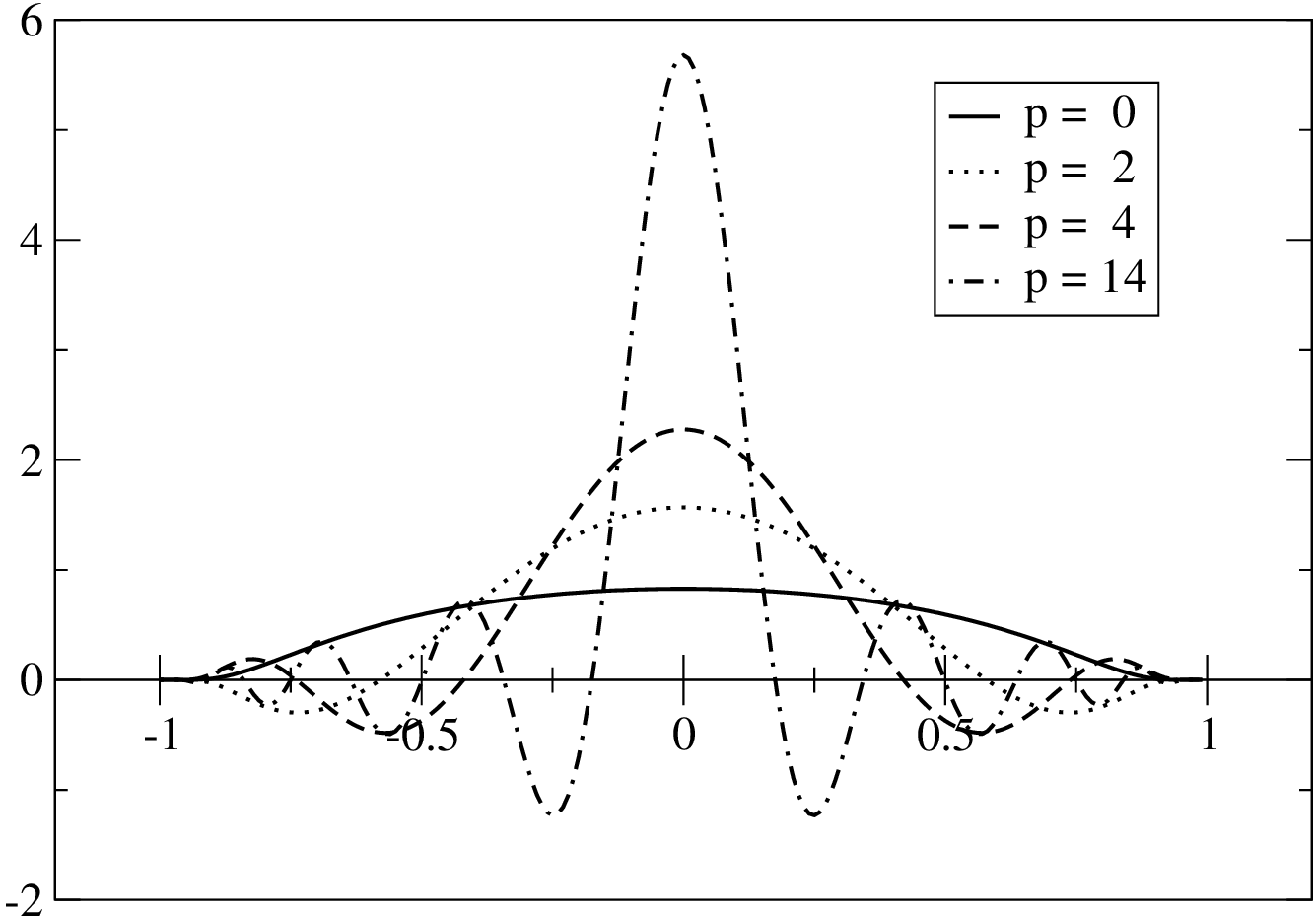}
\caption{Shapes of the finite range smoothing function $f_p(x)$
for $p=0,2,4,14$. Note that $f_0(x)=w(x)$.}
\label{ccps}
\end{figure}

\begin{figure}[tbp]
\includegraphics[width=15.0cm]{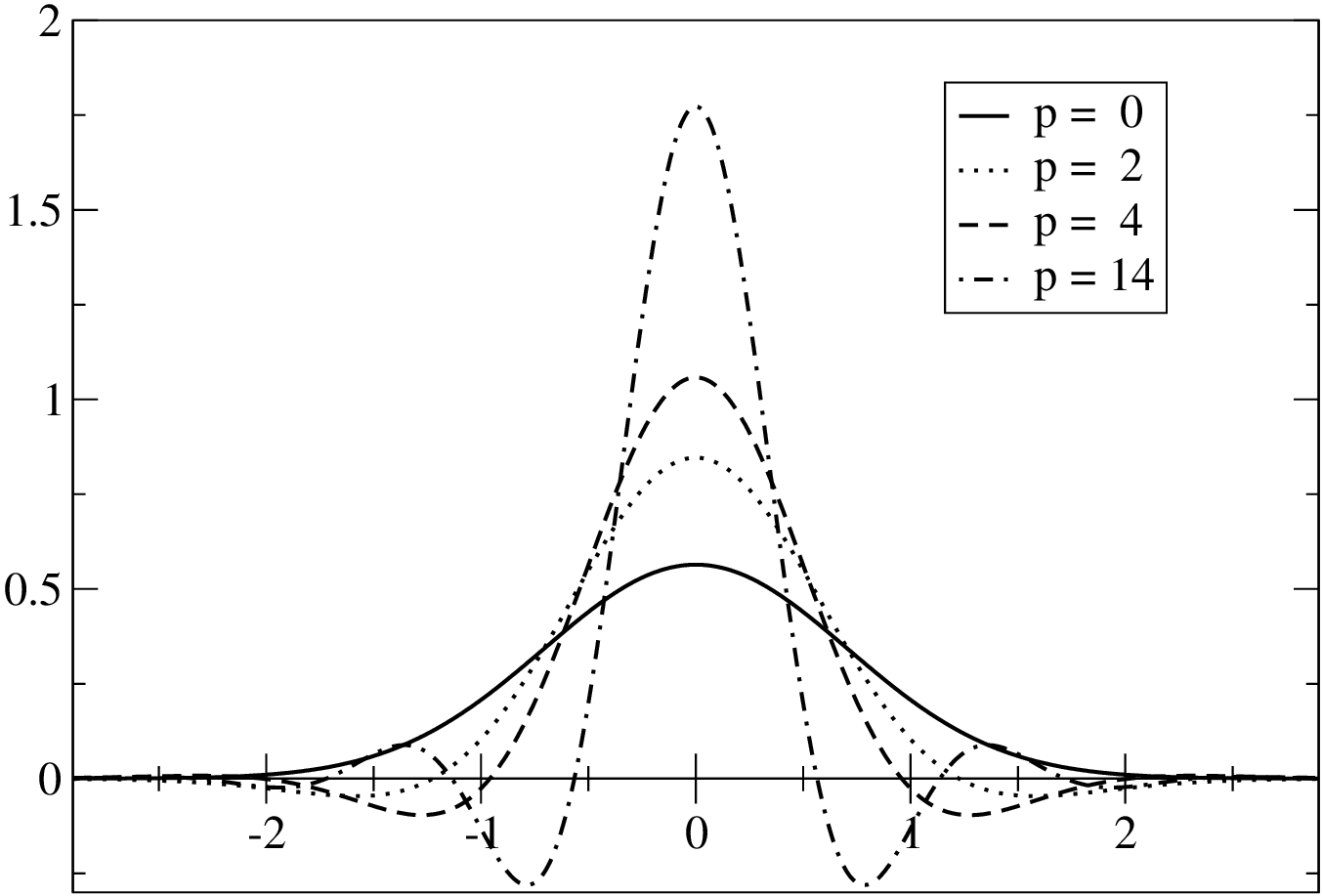}
\caption{Shapes of the smoothing function $f_p(x)$ with Gaussian weight function
for $p=0,2,4,14$. Note that the Gaussian weight function is $f_0(x)=w(x)$. }
\label{ccpsg}
\end{figure}

We present the coefficients $c_i$ for 
the $p\in\{0,2,4,6\}$ values in Table \ref{coeffs} for illustration purposes.
In Fig. \ref{ccps}. we present the shape of the 
smoothing function $f_p(x)$ for a few p values and the finite range weight
function in Eq.(\ref{finitew}) $w(x)=f_0(x)$.
In order to show the difference to the standard Gaussian case, we present
the similar curves with the Gaussian weight function in Fig.\ref{ccpsg}.
For both weight functions for $p>0$
 the
curvature correction polynomials $h_p(x)$ have $p=2m$ zeroes:
\begin{equation}\label{zeros}
h_p(x_j^{(p)})  =0, \quad j=\pm 1 ,...,\pm m, \quad x_{-j}=-x_j~.
\end{equation}
 One can observe the positions of the roots $x_j^{(p)}$ of 
the Eq.(\ref{zeros}) in Fig. \ref{ccps} and Fig.\ref{ccpsg}.
For a fixed $p$ value it is convenient to arrange the positive roots of Eq.(\ref{zeros}) 
so that they form a 
monotonous series:
\be
0<x_1^{(p)}<x_2^{(p)}<...<x_m^{(p)} ~.
\ee
In the smoothing function $f_p(x)$ in Eq.(\ref{smothfunc}) the most
important part of the
smoothing is produced by the central region in $h_p(x)$:
  $x\in [-x_1^{(p)},x_1^{(p)}]$,
determined by the first root $x_1^{(p)}$ . One can see in the figures that
for $p>0$ values 
$x_1^{(p+2)}<x_1^{(p)}$ i.e. the value of $x_1^{(p)}$ decreases when $p$ 
increases.

The finite range smoothing has the advantage that the effect of a certain single particle 
energy $e_i$ vanishes beyond
the interval $E \in [e_i-\gamma,e_i+\gamma]$. Therefore, the
smoothed level density becomes exactly zero for energies lying below
$(e_1-\gamma)$, while the Gaussian oscillates around zero.
This oscillation character appears
at any value of the smoothing parameter. 

If we go to higher $E$-values, we can smooth the oscillatory character of the 
$\tilde{g}(E)$ if we use
large enough $\gamma$ values in the smoothing function with Gaussian
weight function. This is not the case however, if we smooth with finite range
weight function, where 
some undulation in $\tilde{g}(E)$ remains even if we use large
smoothing range parameters.
Therefore, it can not be well approximated by a straight line as it was in the
generalized Strutinski procedure.

This seems to be an important difference between the smoothed level densities 
calculated by using Gaussian or finite range smoothings.

 We calculate the smoothed energy 
in Eq. (\ref{Esmooth}) by using the finite range smoothing functions, for a range of
$\gamma\in[\gamma_{min},\gamma_{max}]$ and
$p\in\{p_{min},p_{min}+2,...,p_{max}\}$ values.
This allows us to study the plateau curves. For $p=0$ the plateau curve is an monotonously
increasing function, therefore, neither the plateau condition in Eq.(\ref{plato})
nor the local plateau condition can be applied. (There is no $\gamma$ value where the derivative is zero.)
This result show the necessity of using curvature correction polynomials.

For $p>0$ plateau curves have minima (and maxima) where the plateau condition in Eq.(\ref{plato})
is fulfilled locally. 
However the plateau curves might have several minima and we have to find the
proper one among those minima. A necessary condition of the smoothing is that the
smoothed level density should not reflect the shell structure of the single
particle levels. Therefore, in the smoothing procedure we have to start searching for 
the minimum of $\delta{E}(\gamma,p)$ from a ($p$-dependent)
$\gamma_{min}$ value with which the shell structure has already disappeared.

The most important characteristics of the single particle spectrum is the
largest gap between the occupied levels.  Therefore, we have to determine the
largest distance between the consecutive occupied levels of the $N$
particles (shell gap) 
\begin{equation}\label{gap}
G  =\max \Big\{(e_{i+1}-e_i)\Big\}~.
\end{equation}
This $G$ value is a more accurate measure of the shell structure of the single particle
energies than the $\hbar \Omega_0$ in Eq.(\ref{hejtav}).
In order to estimate a reasonable $\gamma_{min}$ value,
we have to determine the effective width of the smoothing function with a given
$p$. 
The effective width corresponds to the central peak of $h_p(x)$ in the interval
$x\in[-x_1^{(p)},x_1^{(p)}]$.
Since the effective range of the smoothing function decreases for increasing
$p$, therefore, for larger $p$ value one 
should use larger $\gamma$ values for having the same smoothing effect.
In order to compensate this effect, it is worthwhile to introduce a 
{\em renormalized smoothing range} as follows:
\begin{equation}\label{convers}
\Gamma_p  = x_1^{(p)}\gamma_p~,
\end{equation}
in which the $p$ dependence of the smoothing is considerably reduced.

In order to smooth the fluctuations due to the major shells this $\Gamma_p$ 
range should
be larger than the shell gap 
$\Gamma_p > G$.  To achieve this we introduce a factor $F>1$, and  calculate a 
minimal value for the renormalized range 
$\Gamma_{p,min}=F G$. (We observed that the optimal value for the factor $F$ is $F=1.5-2$ for
light and $F=2.5-3.5$ for heavier nuclei.) Having fixed this minimum we search 
for the first minimum of $\delta{E}(\gamma,p)$ for 
\be
\label{mingam}
\gamma \ge \gamma_{p,min}=\frac{F G}{x_1^{(p)}}.
\ee
This criteria serves as a guide to select the proper minimum of the plateau
curve $\delta{E}(\gamma_p,p)$. For most nuclei the plateau
curves have multiple minima at $\gamma_{p,1}<\gamma_{p,2}<,..,<\gamma_{p,l}$. 
The number of minima $l$ generally increases when $p$ increases. 
We observed that for $p=2$ we have at most two
minima, i.e. $l=1$ or $l=2$ and one of them satisfies the following condition:

\begin{equation}\label{condi}
\Gamma_{2,l}  = x_1^{(2)}\gamma_{2,l}\sim F G~.
\end{equation}
For higher $p$ values the proper minimum should be close to this value since
we reduced the $p$ dependence considerably by using the renormalized smoothing
range.
Therefore,  we have to select the $k$-th minimum, for which
$\Gamma_{p,k}=x_1^{(p)}\gamma_{p,k}\approx\Gamma_{2,l} $~. If we select the smoothing
range according to this criteria then the variation of the corresponding
$\delta{E}(\gamma_{p,k},p)$ values will be small.

\section{Details of the numerical calculations}

We used Saxon-Woods (SW) potential with spin-orbit term. For protons
it was complemented by a Coulomb
potential of uniformly charged sphere with diffuse edge.
(To have this form is necessary for being able to calculate semi-classical
results for comparison.)
The parameters of the potentials were that of the so called {\em universal potential}
given in Ref.\cite{[dud81]}. The depth of the central potential for
neutrons ($\tau=\nu$) $t_3=1/2$ or for protons ($\tau=\pi$) $t_3=-1/2$
\be
V_\tau(Z,N)=-V\Big [1-2 \kappa t_3 \frac{N-Z}{A}\Big ]~,
\ee
where $\kappa=0.86$, $V=49.6$ MeV. The depth of the spin-orbit potential
\be
V_{so}=-\frac{\lambda_{so} V_\tau}{4}\Big (\frac{\hbar}{2\mu c}\Big )^2~,
\ee
with the reduced mass $\mu$ of the nucleon and $\lambda_{so}=35(36)$ for neutrons(protons).
The diffuseness was $a=a_{so}=a_C=0.7$ fm the same for all potential terms.
The radius parameters were $r_0=1.347$ fm, $r_0=r_C=1.275$ fm for neutrons and
protons, respectively, while for the spin-orbit term $r_{so}=1.31(1.32)$ fm 
for neutrons(protons). 
These potential parameters might not be optimal for the individual nuclei but
give a good general $N$, $Z$ dependence all over the nuclear chart at least for
our purpose for testing our method. 

The single particle energies $e_i$ of the single particle Hamiltonian
were calculated by diagonalizing the matrix of the Hamiltonian in h.o.
basis having twenty principal h.o. quanta and maximal orbital angular 
momentum nine. (An increase of the size of the basis did not change the results.)
The same basis was used for diagonalizing the free Hamiltonian (without
nuclear potential terms) to get the positive energies $e_i^{(0)}$ needed to include
the effect of the continuum in the Green's function method described in
Ref.\cite{[Ve00]} in detail. From the difference of the smoothed level densities
of the spectra of the true and the free Hamiltonians the effect of the
artificial nucleon
gas cancels out and we get the same smoothed continuum level density
as we could get by smoothing the continuum level density derived from the
derivative of the scattering phase shifts \cite{[Ve00]}.

\begin{figure}[tbp]
\includegraphics[width=15.0cm]{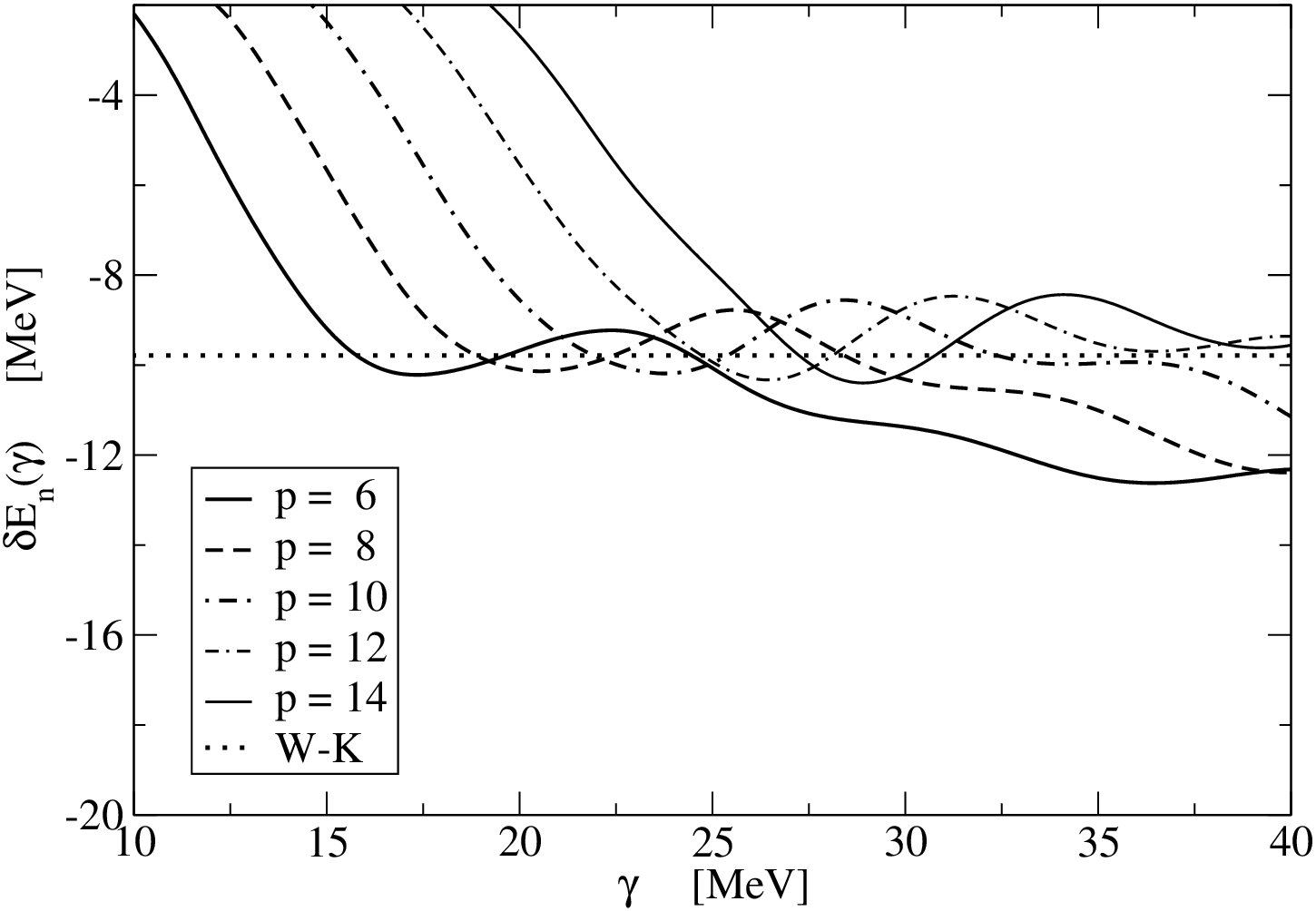}
\caption{Neutron shell corrections $\delta E_n(\gamma,p)$ for the nucleus $^{146}Gd$ as a 
function of the smoothing range
$\gamma$ calculated for $p=6,..,14$ by using the finite-range 
weight function for the smoothing
functions $f_p$. Dotted horizontal line shows the value of
the semi-classical value $\delta E_{sc}=E_{sc}-E_{sp}^n$. }
\label{gdd}
\end{figure}

In Fig. \ref{gdd} we show the {\em plateau curves}  for the
$^{146}Gd$ nucleus with the finite range smoothing and the result of the
Wigner--Kirkwood 
 calculation as a reference.
The range of the $p$ values used in the present work was taken to be the same 
as in
Ref. \cite{[Ve98]} in order to make comparison with those results possible.
Using the new method with the finite range smoothing we are able to use the 
{\em local plateau condition} i.e. to choose the
$\gamma_p$ values where the $\delta {E}(\gamma,p)$ curves have minimum for all
the plateau curves shown. The shell correction values at the minima of the curves
 agree very
well (within 500 keV) with the horizontal line representing the result of the 
semi-classical calculation. Since the $\sigma$ 
variation  of the $\delta {E}(\gamma_p,p)$ values in Eq. (\ref{vari})
is small the shell correction value calculated from the mean in Eq. (\ref{avrage})
is well defined.
 
\begin{figure}[tbp]
\includegraphics[width=15.0cm]{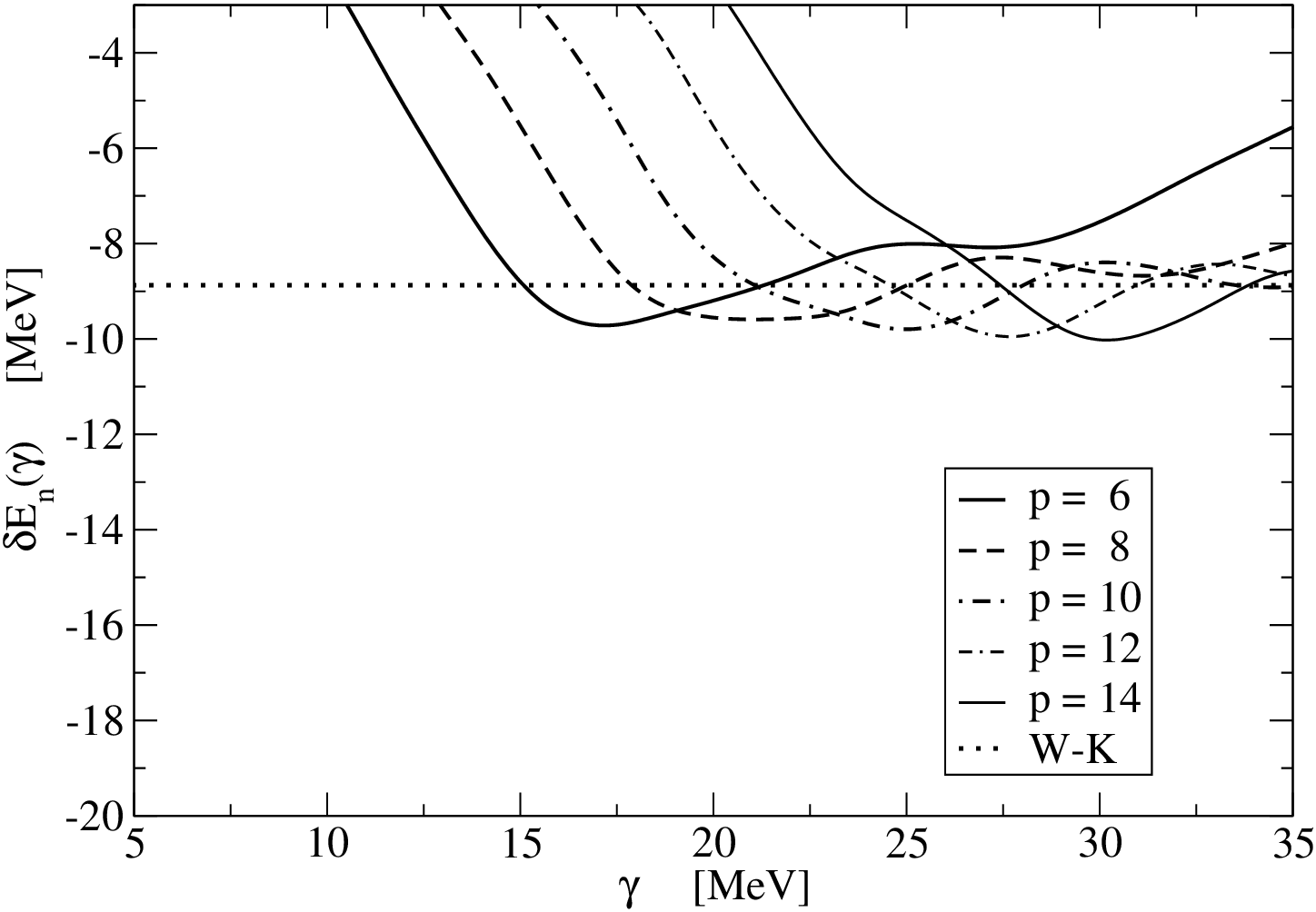}
\caption{Neutron shell corrections $\delta E_n(\gamma,p)$ for the nucleus $^{132}Sn$ as a function of the smoothing range
$\gamma$ calculated for a set of $p$-values by using the finite-range smoothing
function $f_p$. Dotted horizontal line shows the value of
the semi-classical value $\delta E_{sc}=E_{sc}-E_{sp}^n$.}
\label{132sn}
\end{figure}

In Fig.\ref{132sn} we show an example for the double magic $^{132}Sn$
nucleus where the $\sigma$ variation is smaller that 200 keV and the deviation from the
semi-classical value $\Delta$ is less than 1 MeV. This is the largest deviation from the
cases listed in Table \ref{den1}. One can observe in both Figs. \ref{gdd} and
\ref{132sn}, that the $\gamma_p$ values,
where the minima of the $\delta {E}(\gamma_p,p)$
appear are increasing with increasing $p$ values. This can be compensated
to some extent if we use the renormalized smoothing range $\Gamma_p$ defined 
in Eq.(\ref{convers}).

The $\delta{E}(\gamma_p,p)$ plateau curves are very similar for most nuclei we 
calculated if we select the values of the first $\gamma_p$ minima 
of the different $p$ curves beyond $\gamma_{p,min}$ in Eq.(\ref{mingam}).
We identify the shell correction with the mean values of the 
$\delta {E}(\gamma_p,p)$ in Eq. (\ref{avrage}) and its $\sigma$ variation
with the uncertainty of the shell correction.

In Table \ref{den1} 
we show the shell corrections for neutrons and for a set of 
medium and heavy nuclei resulted by the new smoothing procedure 
$\delta E_n(FR)$, and that of the 
generalized Strutinski procedure $\delta E_n(G)$.  Their $\sigma$ variations are
in the third and in the fifth columns. 
In the last two columns we     compare
their values to that of the semi-classical procedure given in Ref.\cite{[Naz94]}.
The differences from $\delta E_{sc}$ are below $1$ MeV for the new procedure
which is a bit better agreement than it is by using the generalized
Strutinski procedure. The average of the differences are $0.6$ MeV and $0.8$
MeV for these two procedure, respectively.

In Table \ref{dep1} 
we show the similar results for protons, where the average of the differences 
from the semi-classical results are $0.4$ MeV and $0.6$
MeV for the new procedure and for the generalized Strutinski procedure, respectively.
So the new procedure can be applied for protons as well.

These differences are not large neither for neutrons nor for protons.
The result of the new procedure is generally closer to the semi-classical 
result if we approach the drip lines. See e.g. the $^{78}Ni$, $^{122}Zr$, 
$^{124}Zr$ nuclei for neutrons and the $^{180}Pb$
nucleus for proton. 
Therefore, we believe that the finite range smoothing allows us to approach the
drip line closer than we can approach it by using the infinite range Gaussian
weight function.

 The basic advantage of the new method is
however, that the determination of the proper shell
correction value is better defined. 
The values resulted by the new procedure are free from most of the
uncertainties of the   generalized
Strutinski smoothing procedure. E.g. they do not depend on the position of the
interval where the linearity of the smoothed level density is required.

The most important advantage of the new procedure is that it can be applied 
for light nuclei where, as we have discussed in Sec.IV.  the
 generalized Strutinski procedure can not be applied.
 
The results of the new method for light nuclei are shown in Table \ref{den} for neutrons and in
Table \ref{dep} for protons. One can see that the agreement with the
semi-classical values are as good  it was for heavier nuclei.
We received specially good agreement for oxygen isotopes even at the neutron
drip line.

\begin{figure}[tbp]
\includegraphics[width=15.0cm]{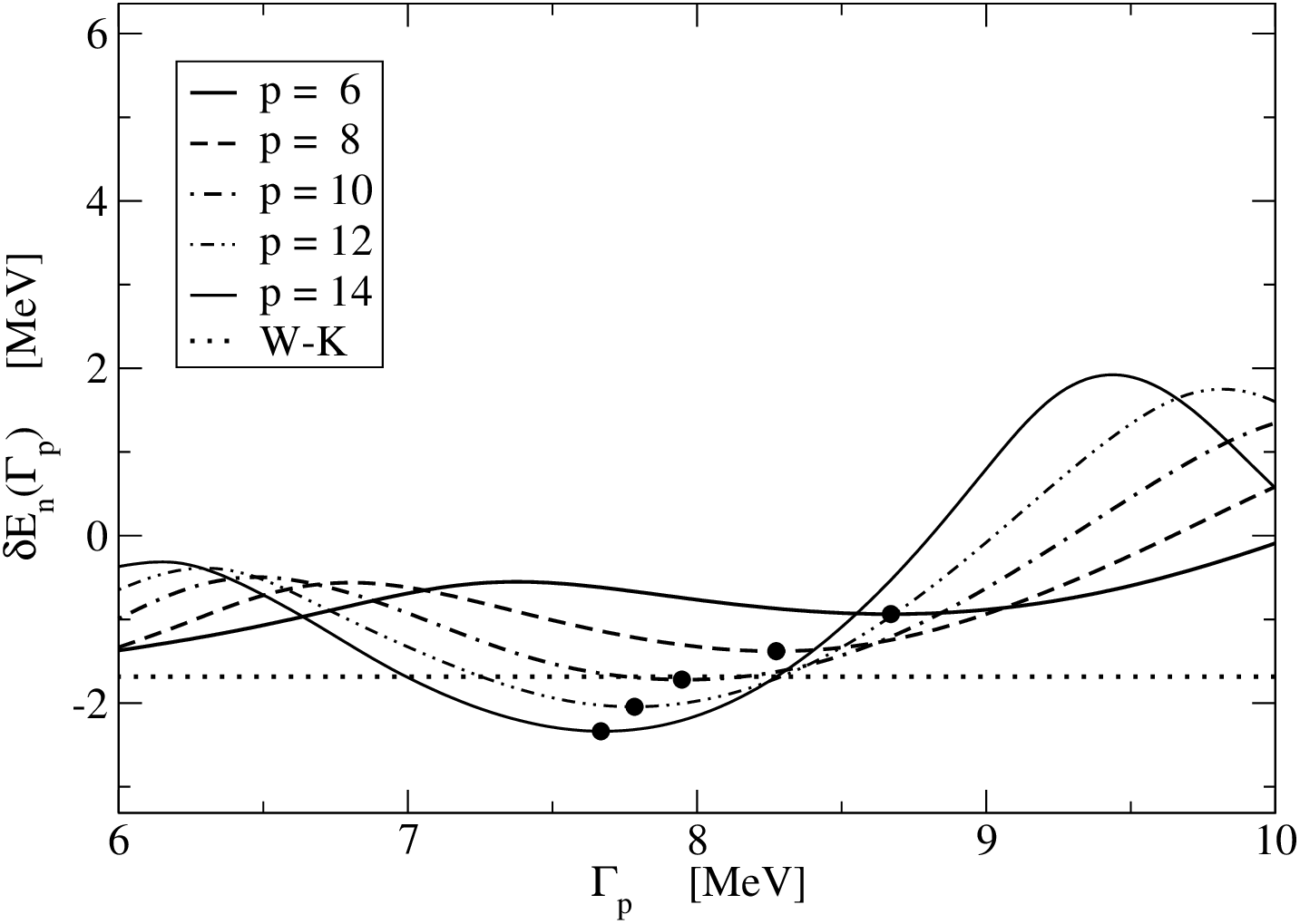}
\caption{Neutron shell correction $\delta E_n(\Gamma_p)$ for the nucleus $^{24}O$ as a function of 
the renormalized smoothing range
$\Gamma_p$ calculated for a set of $p$-values by using the finite-range smoothing
function $f_p$. Dotted horizontal line shows 
the semi-classical value: $\delta E_{sc}=E_{sc}-E_{sp}^n$.}
\label{o24n}
\end{figure}

In Fig.\ref{o24n} we show the neutron plateau curves for the new double magic nucleus
$^{24}O$ as functions of the renormalized smoothing range parameter $\Gamma_p$,
for $p=6,8,...,14$. The semi-classical result is the dotted horizontal line.
The minima of 
each curve  are denoted by filled circles on the 
corresponding curves.
One can see that the $\delta E_n(\Gamma_p,p)$ values denoted by circles are between 
-0.9 and -2.3 MeV and their $\Gamma_p$ values are quite similar at 
$\Gamma_p\sim 8 MeV$. The variation of the $\delta E_n(\Gamma_p,p)$ values are $\sigma\sim 0.5
MeV$ and their mean value coincide with the semi-classical value.
This is certainly an accident but one can see that the $\Delta$ value is
small for the other $O$ isotopes too.
Observe also that the positions of the minima of the different $p$ curves in
this figure scatter much less  in $\Gamma$ ($\sim 15$ \%) than the locations of the minima in
Fig.\ref{132sn} where the smoothing range $\gamma$ was used ($\sim 90$ \%) or in
Fig.\ref{gdd} where the smoothing range $\gamma$ was used ($\sim 70$ \%).
 
Therefore, we believe that the finite range smoothing allows us to approach the
drip line closer than we can approach it by using the infinite range Gaussian
weight function.

\section{Conclusion}         
The new method  uses a finite range smoothing function which makes it possible to localize
the effect of a single particle state with energy $e_i$ to a finite
energy range: $[e_i-\gamma,e_i+\gamma]$. This localization makes possible to extend the
region of applicability of the method to closer to the end regions of the
spectrum. This helps in calculating shell corrections for slightly bound
nuclei lying closer to drip lines and also for lighter nuclei, where the shell
gap is large, therefore, larger values of $\gamma$ values are needed to smooth
the shell structure out.
The new method works equally well for calculating neutron and proton shell
corrections.

 We introduced a renormalized smoothing range in which the $p$
dependence of the smoothing range was reduced considerably. Using this
renormalized range the selection of the proper minimum of the plateau curves
was easier.

 Therefore, we recommend the use of the new procedure with finite range smoothing
first of all for light nuclei, where the generalized Strutinski method
can not be applied. We also recommend its use in regions being close to drip lines
where  the finite range 
smoothing  seems to work somewhat better than the generalized Strutinski method.

\section{Acknowledgement}    
This work has been supported by the 
Hungarian OTKA fund No. K72357.


\begin{table}
\begin{center}
\begin{tabular}{ccccc}
\hline
$p$&$c_0$&$c_2$&$c_4$&$c_6$\\
\hline
$0$&$1$&$0$&$0$&$0$\\
$2$&$1.8934$&$-5.6506$&$0$&$0$\\
$4$&$2.7492$&$-20.62052$&$28.52324$&$0$\\
$6$&$3.5866$&$-48.45461$&$155.33082$&$-136.79695$\\
\hline
\end{tabular}
\end{center}
\caption{\label{coeffs}
Coefficients of the curvature correction polynomials for the lowest
$p$ values corresponding to the finite-range weight function in Eq.(\ref{finitew})}
\end{table}

\begin{table}
\begin{center}
\begin{tabular}{cccccccc}
\hline
Nucleus&$\delta E_n(FR)$&$\sigma$&$\delta E_n(G)$&$\sigma$&$\delta
E_{sc}$&$\Delta_{FR}$&$\Delta_{G}$\\
\hline
\hline
 $~^{ 68}Ni$&$  0.16$&$  0.12$&$  0.50$&$  0.07$&$  0.81$&$  0.65$&$  0.31$\\
 $~^{ 78}Ni$&$ -3.59$&$  0.07$&$ -2.78$&$  0.16$&$ -4.21$&$  0.62$&$  1.43$\\
 $~^{ 90}Zr$&$ -7.42$&$  0.06$&$  -7.35$&$  0.17$&$ -6.82$&$  0.60$&$  0.53$\\
 $~^{ 122}Zr$&$ -5.92$&$  0.11$&$ -4.52$&$   0.15$&$ -6.33$&$  0.41$&$  1.81$\\
 $~^{ 124}Zr$&$ -4.12$&$  0.12$&$ -3.25$&$  0.13$&$ -4.35$&$  0.23$&$  1.10$\\
 $~^{ 100}Sn$&$ -8.16$&$  0.20$&$ -6.95$&$   0.23$&$ -7.50$&$  0.66$&$  0.55$\\
 $~^{ 132}Sn$&$ -9.85$&$  0.14$&$ -8.58$&$  0.10$&$ -8.87$&$  0.98$&$  0.29$\\
 $~^{ 146}Gd$&$ -10.26$&$  0.07$&$ -10.33$&$   0.20$&$ -9.79$&$  0.47$&$  0.54$\\
\hline
\end{tabular}
\end{center}
\caption{\label{den1}
Neutron shell corrections $\delta E_n$ and their variations $\sigma$ calculated using the
finite range weight function $(FR)$ and the generalized Strutinski procedure $G$
in comparison with the
semi-classical shell correction $\delta E_{sc}=E_{sc}-E_{sp}^n$ calculated for
several nuclei. Their deviations from the semi-classical result 
$\Delta_{FR}=|\delta E_{sc}-\delta {E}_n(FR)|$,
\ \ $\Delta_G=|\delta E_{sc}-\delta {E}_n(G)|$ are also shown. All energies are in MeV units.}
\end{table}

\begin{table}
\begin{center}
\begin{tabular}{cccccccc}
\hline
Nucleus&$\delta E_p(FR)$&$\sigma$&$\delta E_p(G)$&$\sigma$&$\delta
E_{sc}$&$\Delta_{FR}$&$\Delta_G$\\
\hline
\hline
 $~^{ 90}Zr$&$  1.59$&$  0.19$&$ 1.88$&$  0.20$&$  1.44$&$  0.15$&$  0.44$\\
 $~^{ 100}Sn$&$ -7.47$&$  0.064$&$-7.42$&$  0.14$&$ -7.01$&$  0.46$&$  0.41$\\
 $~^{ 132}Sn$&$ -7.39$&$  0.068$&$-6.04$&$  0.12$&$ -6.64$&$  0.75$&$  0.60$\\
 $~^{ 146}Gd$&$  4.89$&$  0.10$&$ 5.28$&$  0.24$&$  4.52$&$  0.37$&$  0.76$\\
 $~^{ 180}Pb$&$ -8.94$&$  0.15$&$-7.78$&$  0.04$&$ -8.62$&$  0.32$&$  0.84$\\
 $~^{ 208}Pb$&$ -7.57$&$  0.07$&$-6.73$&$  0.03$&$ -7.29$&$  0.28$&$  0.56$\\
\hline
\end{tabular}
\end{center}
\caption{\label{dep1}
Proton shell corrections $\delta E_p$ and their variations $\sigma$ calculated using the
finite range weight function $(FR)$ and the generalized Strutinski procedure $G$
in comparison with the
semi-classical shell correction $\delta E_{sc}=E_{sc}-E_{sp}^n$ calculated for
several nuclei. Their deviations from the semi-classical result
$\Delta_{FR}=|\delta E_{sc}-\delta {E}_p(FR)|$,
\ \ $\Delta_G=|\delta E_{sc}-\delta {E}_p(G)|$ are also shown.
 All energies are in MeV units.}
\end{table}

\begin{table}
\begin{center}
\begin{tabular}{ccccc}
\hline
Nucleus&$\delta E_n$&$\sigma$&$\delta E_{sc}$&$\Delta$\\
\hline
\hline
 $~^{ 16}O$&$ -1.63$&$  0.04$&$ -1.57$&$  0.06$\\
 $~^{ 18}O$&$  2.67$&$  0.04$&$  3.01$&$  0.34$\\
 $~^{ 20}O$&$  3.25$&$  0.24$&$  3.11$&$  0.14$\\
 $~^{ 22}O$&$  0.12$&$  0.53$&$  0.09$&$  0.03$\\
 $~^{ 24}O$&$ -1.68$&$  0.49$&$ -1.69$&$  0.01$\\
 $~^{ 20}Ne$&$  3.07$&$  0.56$&$  3.01$&$  0.06$\\
 $~^{ 40}Ca$&$ -1.77$&$  0.35$&$ -0.66$&$  0.97$\\
 $~^{ 48}Ca$&$ -2.91$&$  0.24$&$ -2.59$&$  0.32$\\
\hline
\end{tabular}
\end{center}

\caption{\label{den}
Shell correction $\delta E_n$, the variation $\sigma$ 
in Eq.(\ref{vari}) and the
semi-classical shell correction $\delta E_{sc}=E_{sc}-E_{sp}^n$ calculated for
several nuclei. The deviations $\Delta=|E_{sc}-\tilde {E}|$ are also shown. All energies are in MeV units.}
\end{table}

\begin{table}
\begin{center}
\begin{tabular}{ccccc}
\hline
Nucleus&$\delta E_p$&$\sigma$&$\delta E_{sc}$&$\Delta$\\
\hline
\hline
 $~^{ 16}O$&$ -1.65$&$  0.03$&$ -1.44$&$  0.21$\\
 $~^{ 18}O$&$ -1.65$&$  0.10$&$ -1.66$&$  0.01$\\
 $~^{ 20}O$&$ -2.09$&$  0.19$&$ -1.90$&$  0.19$\\
 $~^{ 22}O$&$ -2.30$&$  0.15$&$ -2.14$&$  0.16$\\
 $~^{ 24}O$&$ -3.10$&$  0.66$&$ -2.36$&$  0.74$\\
 $~^{ 40}Ca$&$ -1.62$&$  0.12$&$ -0.91$&$  0.71$\\
 $~^{ 48}Ca$&$ -1.70$&$  0.19$&$ -1.44$&$  0.26$\\
 $~^{ 48}Ni$&$ -0.80$&$  0.36$&$ -1.23$&$  0.43$\\
 $~^{ 56}Ni$&$ -3.67$&$  0.29$&$ -3.45$&$  0.22$\\
\hline
\end{tabular}
\end{center}
\caption{\label{dep}
Shell correction $\delta E_p$, the variation $\sigma$ in Eq.(\ref{vari}) 
and the
semi-classical shell correction $\delta E_{sc}=E_{sc}-E_{sp}^p$ calculated for
several nuclei. All energies are in MeV units.}
\end{table}

\end{document}